\documentclass[twocolumn,twoside,slac_two]{revtex4}
\usepackage{graphicx}
\usepackage{fancyhdr}
\begin{document}

\title{Experimental/Observational Summary - Very High Energy Cosmic
  Rays and their Interactions}

%

\author{Paul Sommers}
\affiliation{Institute for Gravitation and the Cosmos and Penn State Physics,
  University Park, PA 16802, USA}

\begin{abstract}
Observations of cosmic rays have been improving at all energies, with
higher statistics and reduced systematics.  Fundamental questions
remain regarding the origins of cosmic rays both within the Galaxy and
in extragalactic sources, and new puzzles have arisen at ultra-high
energies.  A key issue is determining the elemental composition based
on air shower measurements.  Accelerator experiments at the LHC, with
comprehensive measurements in the forward direction and high
interaction energies, will greatly reduce the uncertainty in air
shower simulations.  Ultra-high energy air showers may reveal
properties of particle interactions at energies far beyond the reach
of the LHC.
\end{abstract}

\maketitle

\thispagestyle{fancy}


\section{INTRODUCTION}

The Sixteenth International Symposium on Very High Energy Cosmic Ray
Interactions has brought together high energy physicists from the
domains of collider experiments and cosmic ray observatories.
Fermilab is a natural setting for this interaction, as it now has
strong participation in both communities.  The instrumentation of high
energy particle physics is common to all.  

Close cooperation and dialog between the communities is now especially
timely, as data from LHC collider experiments begin to establish
particle interaction properties up to higher energy, and measurements
of air showers produced by ultra-high energy cosmic rays offer
constraints on interactions at still higher energies.  Collider
experiments make and measure individual collisions between
particles of known energy and type.  Observations of cosmic ray
interactions are indirect, and the energy and particle type of the
incident cosmic ray must be inferred.  Cosmic accelerators provide an
uncontrolled beam, but that cosmic beam provides access to
interactions with energies far beyond what can be reached in
laboratories.

In his colloquium during the conference, Dietrich Mueller emphasized
the two persistent questions about cosmic rays: What are these
particles?  And where do they come from?  Direct measurements of
cosmic rays at the top of the atmosphere have established the
distribution of nuclear masses at low energies.  These cosmic rays are
known to be of galactic origin from the abundances of radioactive
nuclei, and the column density (g/cm$^2$) of matter which they
traverse in the disk of the Galaxy before escaping is determined for
different energies from the relative abundance of isotopes that are
produced by spallation.  GeV and Tev gamma ray observations provide
evidence supporting the hypothesis that supernovae and pulsar wind
nebulae produce these cosmic rays in the Galaxy, although observational
proof is still lacking that those gamma rays are produced by cosmic
ray interactions rather than electromagnetic processes.  

Direct measurements of electrons and positrons (ATIC and PAMELA) have
found excesses that have prompted widespread speculation that they
could be evidence for dark matter annihilation or decay.
Alternatively, the overdensities relative to expectation might be due to
one or more nearby sources.  Curious anisotropy features observed both
by Milagro and ARGO might also be evidence of cosmic ray particles
from nearby sources.

The answers to Mueller's questions are even less clear at energies where
the low flux requires indirect measurements by air showers.  This
includes the especially interesting region of the spectrum's knee near
3 PeV.  Open questions remain as to what causes the knee.  Is it a
rigidity limit of supernova shock acceleration, with the spectrum for
each nuclear type breaking at an energy proportional to its charge?
Does the knee result from an abrupt change in the rigidity dependence
of the time needed for escape from the Galaxy?  Or could it be a
measurement artifact stemming from an interaction energy threshold
effect that would create a spectral break at a particular value of E/A
rather than E/Z?  Is the sharpness of the knee indicative of a single
prominent source?

Of particular interest is the energy of transition where the low
energy cosmic rays of galactic origin give way to a different
population that come from extragalactic sources.  A transition between
two power law spectra would necessarily be concave upward in the
transition region (on a log-log plot of the spectrum).  The only
prominent upward concavity in the spectrum is at the ankle near 5 EeV.
(1 EeV = $10^{18}$ eV.)  Dominance by galactic cosmic rays to such
high energy requires something other than supernova accelerators,
i.e. the galactic ``source B'' of Hillas (cf. Gaisser's talk
\cite{Gaisser}).  Another popular picture is that the ankle is a
feature carved from an extragalactic proton spectrum by $e^{\pm}$ pair
production, with the transition energy being somewhere below the ankle
\cite{Berezinsky}.  In that case, the challenge is to find evidence
for the transition in the energy spectrum or composition at some
energy below the ankle.  KASCADE Grande extended upward the energy
range of the KASCADE array for this purpose.  The TALE extension of
the Telescope Array and the HEAT enhancement of Auger are extending
downward their energy ranges for this search.  Evidence for upward
concavity just before a ``second knee'' was shown in talks by
Arteaga-Velázquez \cite{Arteaga} and Martirosov \cite{Martirosov}.
See section \ref{Below-Ankle}.

Ironically, the answers to Mueller's questions may be easier for the
extremely rare trans-GZK \cite{GZK} particles with energy above
$6\times 10^{19}$ eV.  Cosmic rays cannot retain such high energy for more
than roughly 100 Mpc due to pion photoproduction (protons) or nuclear
photodisintegration.  Their sources must therefore lie within that
``GZK sphere,'' and protons can arrive from those sources with
magnetic deflections of only a few degrees.  Also, except for heavy
nuclei like iron, photodisintegration is so rapid that any
contributing sources of nuclei must be within just tens of Mpc.
Unless there are one or more strong sources very close, the
composition can only be protons, heavy nuclei like iron, or some
mixture of just those two types.  Heavy nuclei are deflected much more
than protons by magnetic fields (both galactic and extragalactic).
Compelling evidence for small deflections from candidate sources would
identify the sources and also establish a population of extremely
high energy protons.  That leads to the next questions: How are the
protons accelerated in those sources?  And what do we learn about high
energy hadronic interactions from air shower development properties
using that proton beam?  Intriguing results from the Auger Observatory
are suggesting some exciting results in this direction.  Analyses of
HiRes data, however, do not confirm trans-GZK anisotropy, and there
are differences in measured properties of air shower developments
between HiRes and Auger.

There were a large number of exciting talks about experiments and
observations -- too many to summarize here.  The written versions 
are available in these proceedings.  This summary is an eclectic
selection of topics and results from the presentations.  

\section{Accelerator experiments}

The exciting fact is that the LHC is operational.  Proton collisions
are occurring at 7 TeV center-of-mass energy.  Detectors are
collecting lots of data and results are being published.  

Mike Albrow \cite{Albrow} gave a helpful introduction to accelerator
data for purposes of cosmic ray physics with a historical perspective.
He focused on hadronic collision results in the forward region above
about 20 GeV center-of-mass energy.

Rajendran Raja \cite{Raja} explained the importance of the MIPP
upgrade for cosmic ray physics.  It will study interactions of six
different beam particles (protons, kaons, and pions) on a large number
of nuclei, with full acceptance over phase space, including nuclear
fragmentation.  The plan is to change the target nucleus each day and
collect about 5 million events in a day.

Baha Balantekin \cite{Baha} summarized results from heavy ion
collisions at RHIC that should be relevant for cosmic ray nuclei
interacting with atmospheric nuclei.  The quark-gluon state is almost
a perfect fluid, and RHIC has measured its temperature in gold-gold
collisions.

Mary Convery \cite {Convery} reviewed recent results from D0 and CDF
at the Tevatron.  These include single top measurements, new heavy
baryons, a possible signature of CP violation beyond the standard
model in di-muon charge asymmetry, and constraints on the Higgs mass.

Switching to LHC, Georges Azuelos \cite{Azuelos} reported on ATLAS.
It was tested on cosmic rays and is collecting quality LHC data,
validating the detector simulations.  It is poised to measure cross
sections, efficiencies, and rare processes, and to look for unexpected
phenomena.

Ambra Gresele \cite{Gresele} presented early results from CMS.  Papers
have reported rapidity and transverse momentum distributions at
several energies, including 7 TeV and Bose-Einstein correlations.
Single-diffractive events have been observed in the calorimeters.

TOTEM is potentially one of the most important detector systems for
cosmic rays as it is designed to measure the total cross section and
forward charged particle multiplicity distributions.  Emilio Radicioni
\cite{Radicioni} talked about its construction and readiness.  It is
still in a commissioning state until after the winter shutdown.

In conjunction with ATLAS, LHCf studies very forward neutral
particles.  Takashi Sako \cite{Sako} reported on the performance of LHCf and
preliminary results at 900 GeV and 7 TeV.  They have almost enough
statistics and will focus on systematics before finalizing results.
The detectors will be removed for radiation hardening ahead of the 14
TeV runs.  

At 14.4 meters from the CMS interaction point, CASTOR is a Cherenkov
calorimeter that surrounds the beam pipe and is sensitive to forward
particles -6.6$<\eta<$-5.2.  Edwin Norbeck \cite{Norbeck} presented
its status and performance, emphasizing searches for exotic events
like cosmic ray centauros, strangelets, and long penetrating
particles.

Christian Linn \cite{Linn} reported on early performance and results
from LHCb.  It specializes in precision measurements of B decays.  The
K$_S$ differential production cross section is slightly harder than in
Monte Carlo models.  $\bar{\Lambda}/\Lambda$ production ratio is lower
than MC tunings at 900 GeV, but agreement with predictions is good at
7 TeV.

\begin{figure}[ht]
\includegraphics[width=80mm]{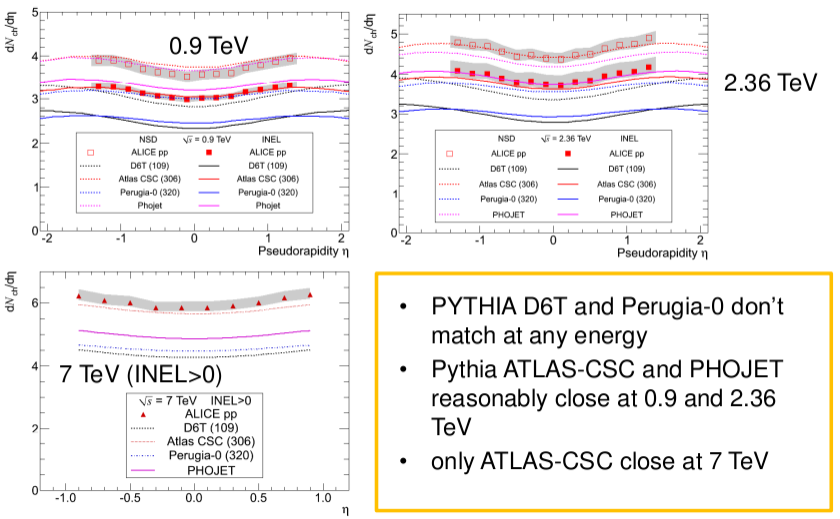}
\caption{ALICE results for dN$_{ch}$/d$\eta$ and comparison to model
  predictions.} \label{ALICE}
\end{figure}

Results from ALICE were reported by Henner Buesching \cite{Buesching}.
These include multiplicity distributions at 900 GeV, 2.6 and 7 TeV,
transverse momentum distribution at 900 GeV and mean p$_T$ as a
function of multiplicity, and the anti-baryon/baryon ratio at 900 GeV
and 7 TeV.  Figure \ref{ALICE} shows a comparison of data with some
model expectations.
Although PYTHIA does not provide a good fit to the data, the
models used in cosmic ray physics comfortably bracket the data, and
Sibyll 2.1 gives good agreement, as shown in the talk of Tanguy Pierog.

\section{Muon measurements with laboratory detectors}

Detectors built for studies of particle collisions and neutrino beams
have been used also to detect cosmic ray muons.  Those muons are
helpful in analyzing detector performance, but the measurements can
also be used for cosmic ray physics.

Yuqian Ma \cite{Ma-1} reported on results from the L3+C collaboration.
They have measured the atmospheric muon energy spectrum, the muon
charge ratio; the moon's shadow, the anti-proton/proton ratio, and
properties of muon bundles.  They have not found any muon excess from
candidate point sources, but they identified a hot spot as a candidate
unknown source.  They do not detect anisotropy in sidereal time.

CMS results for muon measurements were reported by Gavin Hesketh
\cite{Hesketh}.  Measurements were made above ground and in the cavern
underground.  The charge ratio has been measured carefully as a
function of energy, and results are consistent with cosmic ray shower
models \cite{CMS_muons}.

Philip Schreiner \cite{Schreiner} reported results from MINOS on
atmospheric muons.  They make careful corrections for temperature,
which affects their muon detection rates.  They have been able to
determine several meson production rate ratios for primary cosmic rays
above 7 TeV: $\pi^+/\pi^-$, $K^+/K^-$, and also
$(K^++K^-)/(\pi^++\pi^-)$.

\section{Direct cosmic ray measurements}

The direct measurements were divided according to whether or not the
detectors used magnetic spectrometers.

John Mitchell \cite{Mitchell} reviewed the missions with detectors
that include magnetic spectrometers.  Those include both missions in
space as well as balloon payloads.  The magnet allows charge
separation for particles of the same mass and energy, so in
conjunction with other detectors these can distinguish matter from
anti-matter.  The talk focused primarily on four different missions:
(1) the two balloon flights of Bess-Polar (Balloon-Borne Experiment
with a Superconducting Spectrometer), (2) PAMELA satellite (Payload
for Antimatter Matter Exploration and Light-nuclei Astrophysics), (3)
AMS (Alpha Magnetic Spectrometer), its first ride on the space station
and plans for the upcoming AMS-02, (4) the future PEBS (Positron
Electron Balloon Spectrometer).

\begin{figure}[ht]
\includegraphics[width=60mm]{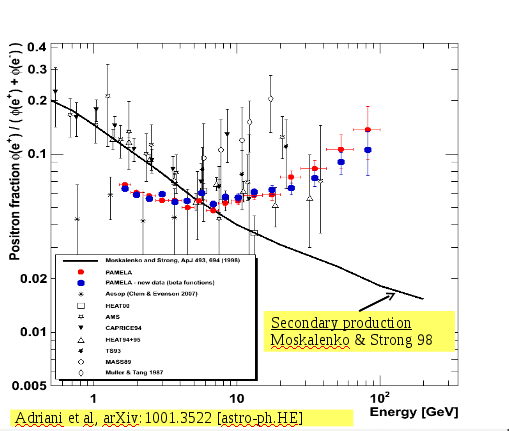}
\caption{Positron to electron fraction measured by PAMELA as a
  function of energy.  This figure includes data taken through the end
of 2008.  } \label{PAMELA}
\end{figure}

Results on the energy dependence of the positron/electron ratio from
PAMELA have attracted enormous attention due to possible explanations
in terms of dark matter annihilations.  The updated PAMELA plot is
shown in Figure \ref{PAMELA}.  Mitchell's talk included not only a
discussion of various dark matter scenarios, but also astrophysical
explanations in terms of a nearby source (supernova or pulsar).

Andrei Kounine \cite{Kounine} presented details about the AMS mission.
In particular, he explained the reason for the recent announcement
that the permanent magnet used in AMS-01 would be used also in AMS-02
instead of the planned superconducting magnet.  The hope is to collect
data until 2020 or 2028.  With the superconducting magnet, the
lifetime would be limited to less than 3 years by the cryogenics,
without an option to refill.  The detector has been reconfigured to
work with the weaker field of the permanent magnet, and launch is
expected in November of this year.

Eun-Suk Seo \cite{Seo} reviewed the missions that measure cosmic
rays directly without a magnetic spectrometer.  These include ATIC
(Advanced Thin Ionization Calorimeter), Fermi Gamma-ray Space
Telescope, CALET (Calorimetric Electron Telescope), CREST (Cosmic Ray
Electron-Synchrotron Telescope), TRACER (Transition Radiation Array
for Cosmic Energetic Radiation), TIGER (Trans-Iron Galactic Element
Recorder), and CREAM (Cosmic Ray Energetics and Mass).  The ATIC
collaboration reported a significant excess electron intensity near
500 GeV in 2008 (relative to model expectations), which, taken along
with the PAMELA result, ignited interest in the interpretation of dark
matter annihilating to a light boson.  Astrophysically the excess flux
can be attributed to the presence of individual sources \cite{Profumo}

\begin{figure}[ht]
\includegraphics[width=60mm]{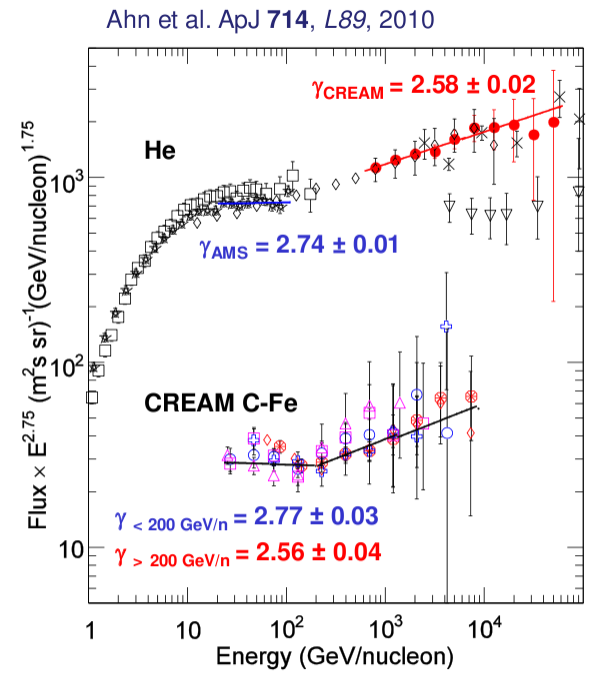}
\caption{CREAM energy spectra showing an upward concavity near 200 GeV.} \label{CREAM}
\end{figure}

Seo emphasized results from the five flights of CREAM: the
boron/carbon ratio up to TeV/nucleon, and elemental spectra over four
decades of energy.  The energy spectra show a distinct hardening near
200 GeV/nucleon (Figure \ref{CREAM}), and a variety of explanations
have been proposed to account for it.  

\begin{figure}[ht]
\includegraphics[width=60mm]{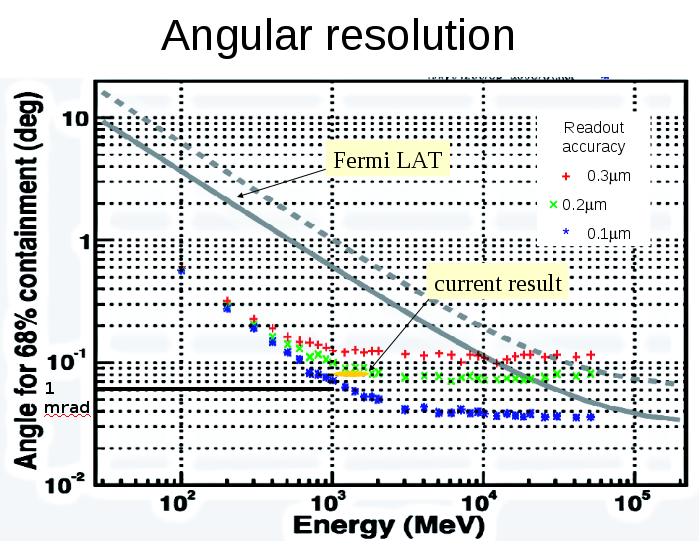}
\caption{Angular resolution of the balloon-borne nuclear emulsions for
gamma rays.} \label{Takahashi}
\end{figure}

Satoru Takahashi \cite{Takahashi} presented plans for a balloon-borne
gamma-ray telescope with nuclear emulsions.  It is expected to have
excellent angular resolution (Figure \ref{Takahashi}) and polarization
sensitivity.

\section{Air shower measurements below the ankle}\label{Below-Ankle}

\begin{figure}[ht]
\includegraphics[width=60mm]{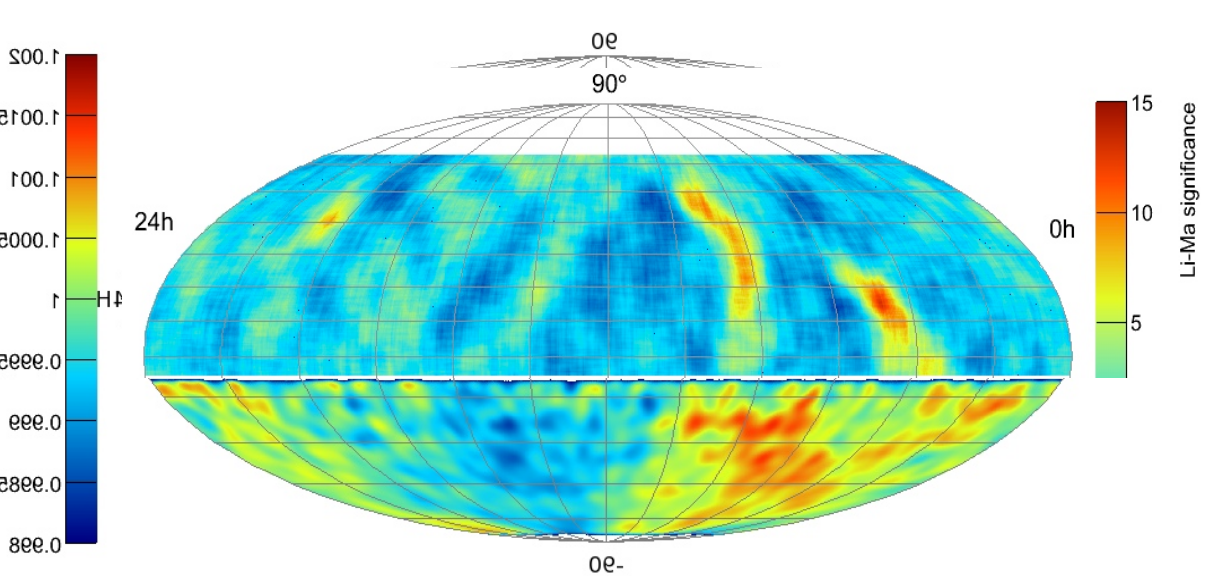}
\caption{Color-coded map of celestial excess and deficit regions.  The
northern hemisphere is derived from Milagro data, and the southern sky
from IceCube.} \label{goodman-A}
\end{figure}

\begin{figure}[ht]
\includegraphics[width=60mm]{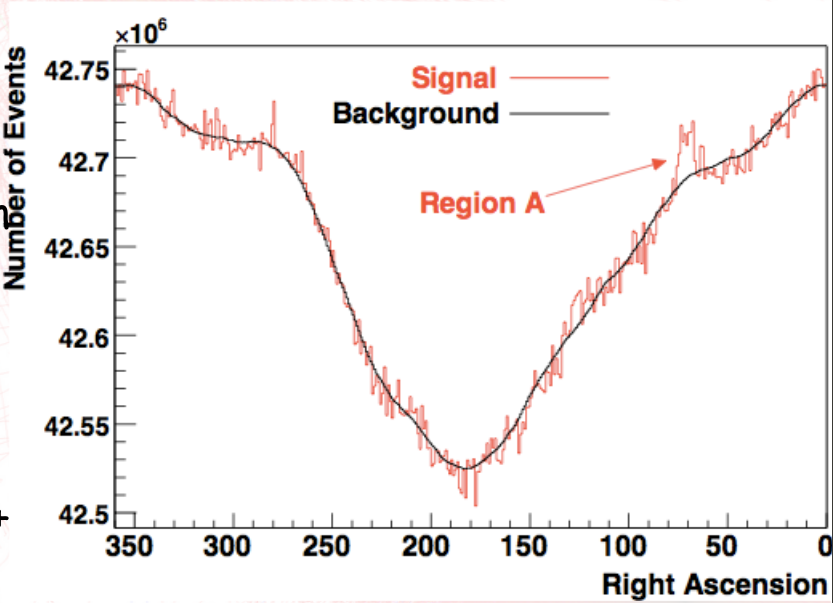}
\caption{Milagro cosmic ray intensity variation with right ascension
  for declination between 10 and 20 degrees.} \label{goodman-B}
\end{figure}

\begin{figure}[ht]
\includegraphics[width=60mm]{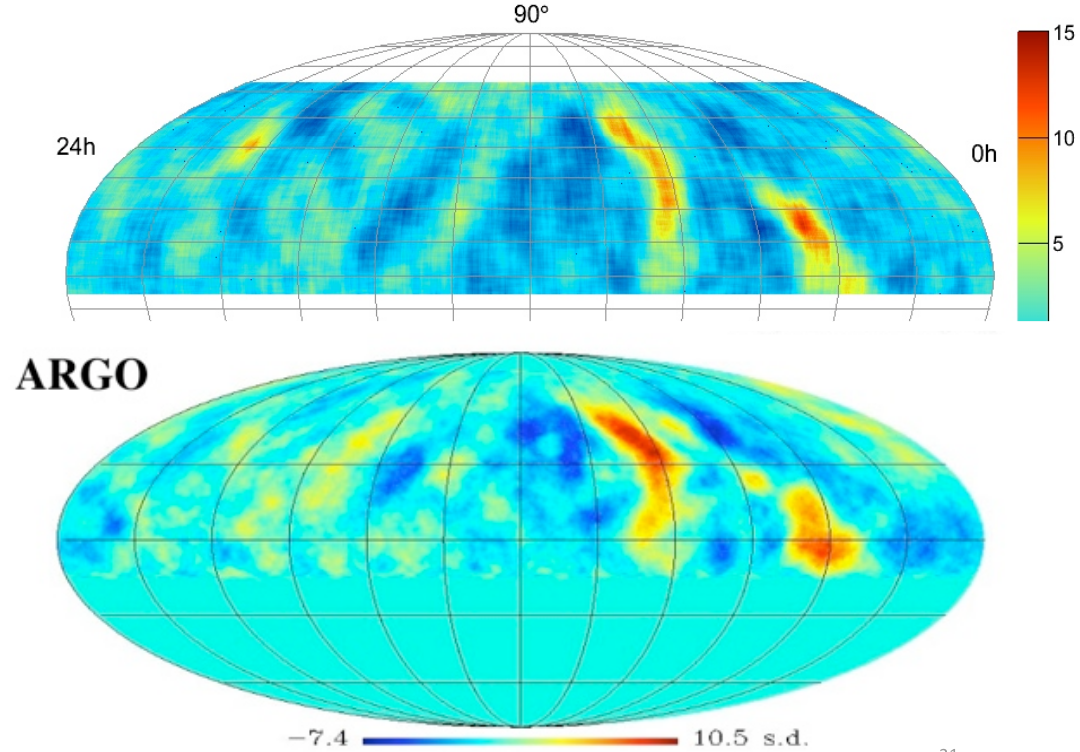}
\caption{Milagro (top) and ARGO data show the same two narrow features
(regions A and B) of excess cosmic ray arrivals.} \label{goodman-C}
\end{figure}

Although direct measurements of cosmic rays have provided rich details
about the energy spectra of individual components up to energies
exceeding 100 TeV, important additional information about anisotropy
of those cosmic rays is coming from air shower detectors and large muon
detectors.  In particular, Jordan Goodman \cite{Goodman} summarized
anisotropy results from Milagro, IceCube, ARGO, and the Tibet Air
Shower Array that show a consistent large scale anisotropy pattern.
Figure \ref{goodman-A} shows a full-sky anisotropy map using IceCube
(muon) data for the southern sky and Milagro cosmic ray data for the
north.  
A one-dimensional plot of the intensity variation with right ascension
is shown in Figure \ref{goodman-B} for a strip of declinations between
10$^{\circ}$ and 20$^{\circ}$.  
The large-scale variation with right
ascension is statistically unquestionable.  The amplitude ($0.1\%$ is
less than the Compton-Getting \cite{Compton-Getting} effect ($0.5\%$)
which would be expected if our motion relative to the cosmic rays were
that of our motion relative to the CMB, and the dipole component is
not in the direction of that motion.  It is interesting that Milagro
has measured a modest time dependence of the anisotropy that is not
confirmed by Tibet AS.  Figure \ref{goodman-B} shows a narrow feature
marked as ``Region A'' in that declination strip.  It is seen, along
with ``Region B'' in Figure \ref{goodman-C}, where the same narrow
features are clear also in the ARGO data.  
These features are due to cosmic rays, not gamma rays, in the Milagro
data, and it is a fascinating open question how charged particles with
such small Larmor orbits in the galactic magnetic field can produce
these narrow features on the sky.

\begin{figure}[ht]
\includegraphics[width=60mm]{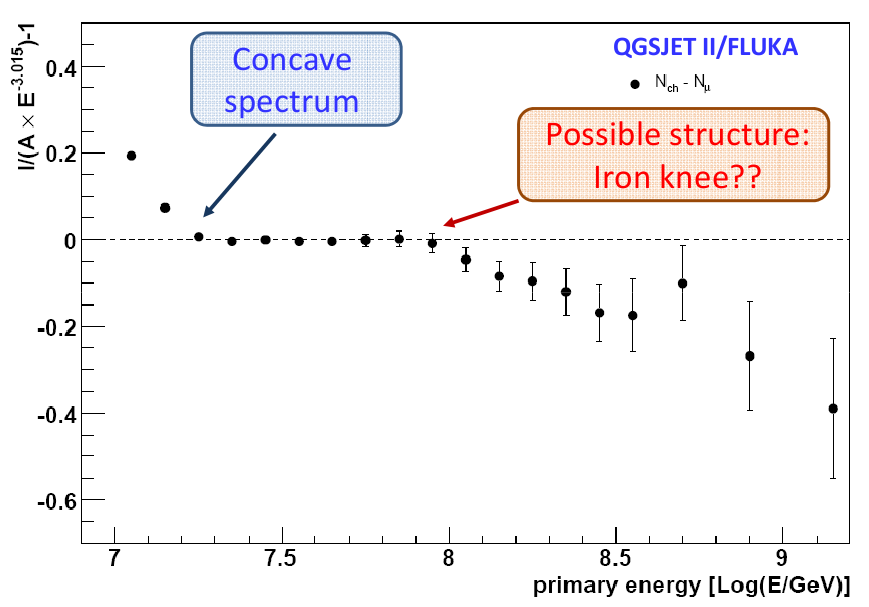}
\caption{The KASCADE-Grande all-particle spectrum is not described by
  a single power law.} \label{Arteaga}
\end{figure}

Results above the knee of the spectrum were reported from
KASCADE-Grande by J.C. Arteaga-Velázquez \cite{Arteaga}.  Inferences about the
composition based on separate measurements of muons and
electromagnetic particles have dependence on which hadronic
interaction model is used (QGSJET or Sibyll).  The all-particle
spectrum cannot be fit by a simple power law above the knee.  Figure
\ref{Arteaga} shows a region of upward concavity followed by a
steepening that might be interpreted as an ``iron knee.''  The upward
concavity was not proposed as a transition to extragalactic cosmic
rays.  Like the spectral hardening seen at lower energy by CREAM,
favored interpretations are shock broadening by the accelerating
particles themselves \cite{Ellison} or the effect of a nearby source.

The KASCADE collaboration is pursuing multiple avenues for determining
the elemental composition of the cosmic rays.  One particularly
interesting approach is muon tracking, as presented by Paul Doll
\cite{Doll}.  The idea is to use the muon arrival directions to
determine the heights of muon production and the muon pseudorapidity
distribution.

\begin{figure}[ht]
\includegraphics[width=60mm]{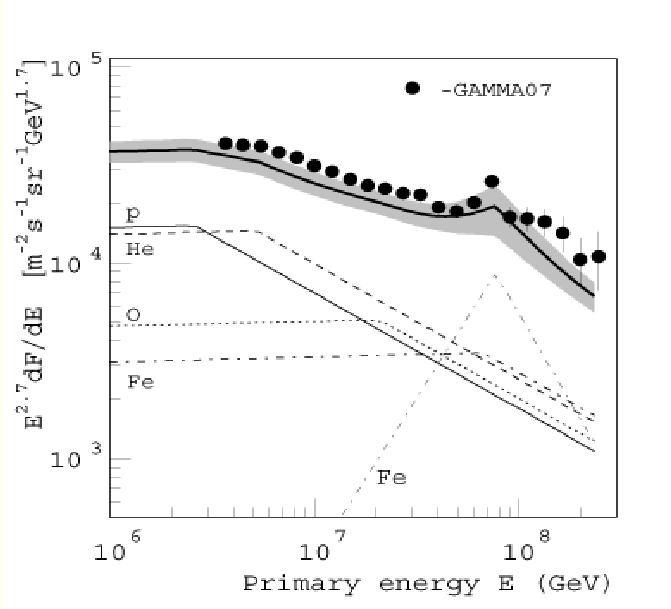}
\caption{Data from Mt. Aragats also deviates from a simple power law
  with a bump-like feature that includes a narrow region of upward
  concavity.} \label{Martirosov}
\end{figure}

As reported by Romen Martirosov \cite{Martirosov}, the GAMMA
experiment on Mt. Aragats has also found an upward concavity followed
by a steepening of the energy spectrum just below $10^{17}$ eV.  The
bump, shown in Figure \ref{Martirosov} may be consistent with the
KASCADE features of Figure \ref{Arteaga}.

IceTop is an air shower array built directly above the IceCube
Observatory at the South Pole, with a pair of ice Cherenkov tanks
deployed above each IceCube string.  Serap Tilav \cite{Tilav}
presented the status of the array as IceCube nears the end of its
construction.  The high elevation is favorable for detecting air
showers near maximum development with an energy threshold near 100
TeV, and the ability to study high energy muon bundles with IceCube
will be a powerful tool for analyzing the composition in the region of
the knee.  The build-up of snow on older detectors affects the trigger
rate and signal levels and must be considered in the analysis.  A raw
energy distribution shows the spectral steepening at the knee.

Sunil Gupta \cite{Gupta} presented results from the GRAPES-3 detector
at Ooty.  It also has a low energy threshold and can compare its
composition analyses based on air showers against direct measurements.
They have measured their angular resolution using the shadow of the
moon.  They have studied in detail the trigger rate dependence on
atmospheric pressure and temperature.  In addition, the rates respond
to space weather, and they track the effects of coronal mass
ejections.

A presentation by Jing Huang \cite{Huang} for the Tibet Air Shower
$\gamma$ Collaboration was given by Yuqian Ma.  The results focused on
the spectrum and composition at the knee.  The Tibet results indicate
that the knee is produced by nuclei heavier than helium, as they
dominate the lightest elements by that energy.  Beyond the knee the
composition is expected to be heavy.  To check these results in
detail, they plan to construct, in three stages, the YAC: Yangbajing
Air shower Core array.  Its goal is to study spectra of individual
components at the knee.  YAC-I is already operational, and some early
results were shown.

\section{Ultra-high energy cosmic rays}

The study of the highest energy cosmic rays is challenged by the
extremely small flux of particles.  The Auger Observatory in Argentina
runs continuously with an aperture of 7000 km$^2$sr.  Despite that
large size, the results on anisotropy and composition are limited
primarily by inadequate statistics.  Paolo Privitera described plans for
Auger North, which would operate with 47,000 km$^2$sr, increasing the
aperture by almost a factor of eight, while retaining good control of
systematics with hybrid detections at night and careful atmospheric
monitoring.  

Jim Adams \cite{Adams} presented plans for JEM-EUSO, a space-based air
fluorescence telescope to be flown on the International Space Station.
Expected to launch in 2015, it will have an enormous aperture by
virtue of a wide field of view (60$^\circ$) and its large distance
from the air showers.  Its duty cycle is limited by sunlight,
moonlight, high clouds, and city lights.  Its energy threshold will be
approximately 70 EeV, roughly the energy above which Auger has
reported cosmic ray anisotropy.  It will have full sky coverage, and
it is expected to distinguish neutrinos, gamma rays, and hadronic
cosmic rays.

Chris Williams \cite{Williams} spoke about MIDAS, an experimental
study of the feasibility of observing air shower longitudinal
developments day and night by using molecular bremsstrahlung
radiation instead of air fluorescence.  Overcoming the limited duty
cycle of air fluorescence telescopes would be a major advance for the
study of ultra-high energy cosmic rays.

Fred Kuehn \cite{Kuehn} reported on the status of the AirFly
measurements of the air fluorescence yield.  That yield is an
important normalizing constant for air fluorescence detectors that is
the basis of the energy scale for ultra-high energy cosmic ray
observatories.  Relative yield dependence on atmospheric pressure,
temperature, and humidity, on photon wavelength and electron energy
have all been published already.  A final result for the absolute
yield is expected by the end of 2010 with uncertainty less than 5\%.

Masaki Fukushima \cite{Fukushima} presented preliminary results
obtained with the Telescope Array in Utah, including a hybrid energy
spectrum and composition analysis using stereoscopic measurements.
The results are consistent with published HiRes results.  In
particular, the mean depths of maximum $X_{max}$ in three energy bins
below and above 10 EeV are at least as deep as expected for protons
using conventional air shower models.  Of special interest is an
upcoming end-to-end calibration of their fluorescence detector using a
linac to accelerate electrons in an upward-going beam at a distance of
100 meters from a TA telescope.

TALE is a proposed low-energy extension of the TA which would permit
air fluorescence measurements of cosmic ray shower longitudinal
profiles down to energies below 100 PeV.  Charlie Jui \cite{Jui}
presented plans for 15 additional telescopes to cover an elevation
angle range from $31^\circ$ to $73^\circ$ over a $90^\circ$ azimuthal
range.  At those lower energies, air showers can only be measured
relatively nearby, and the depth of maximum is therefore viewed at a
high elevation angle. 

\begin{figure}[ht]
\includegraphics[width=60mm]{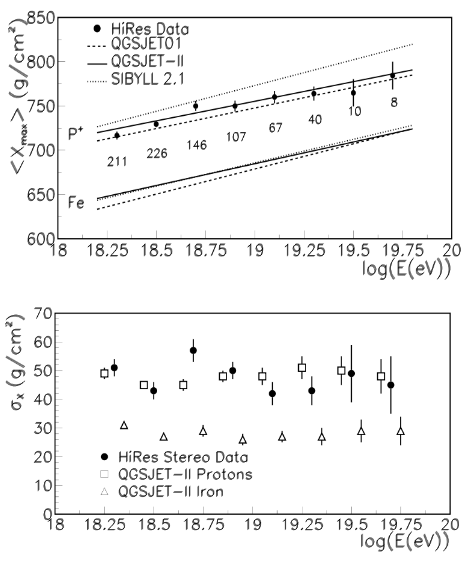}
\caption{HiRes results on air shower depths of maximum $X_{max}$.  The
upper plot shows the mean $X_{max}$ as a function of energy.  The
lower plot shows the width of the $X_{max}$ distribution at each
energy.  In both plots the data are compared with expectations from
customary models for protons and for iron.} \label{hires_xmax}
\end{figure}

Pierre Sokolsky \cite{Sokolsky} reported ``Final Results from the High
Resolution Fly's Eye (HiRes) Experiment.''  Those results include the
stereo energy spectrum, anisotropy, and composition analysis.  The
energy spectrum was shown to agree well with the Auger energy spectrum
by a shift of energies commensurate with the combined systematic
uncertainty of the two observatories.  No significant anisotropy is
found - not the clustering reported by AGASA, not the AGN correlation
reported by Auger, not any detected correlation with nearby large
scale structure represented in the 2MRS catalog, and no confirmation
of a tentative HiRes correlation with BL Lacs reported earlier.  The
HiRes analysis of depths of maximum is shown in Figure
\ref{hires_xmax}.  The results are consistent with simulations of
proton-initiated air showers.  

\begin{figure}[ht]
\includegraphics[width=80mm]{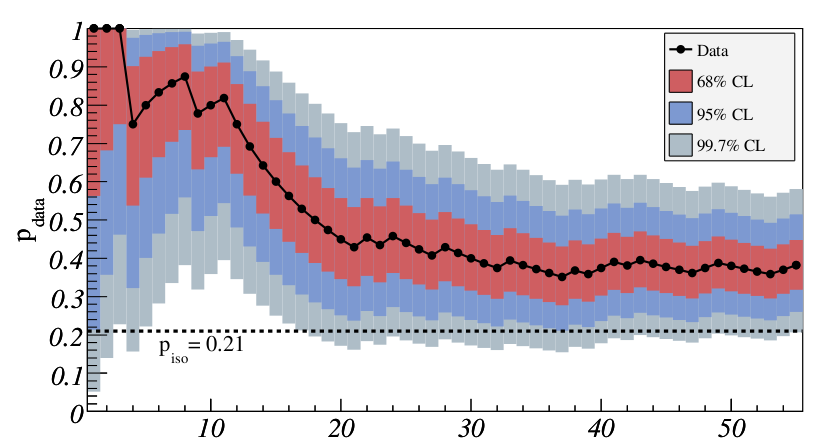}
\caption{Fraction of correlating Auger events above 55 EeV as a
  function of the cumulative number collected {\it after} the
  exploratory scan.  As originally prescribed, an arrival direction
  correlates if it is less than 3.1 degrees from one of the AGNs with
  redshift less than 0.018 in the 12th Veron-Cetty and Veron AGN
  catalog.} \label{auger_corr}
\end{figure}

Paolo Privitera \cite{Privitera} presented results from the Pierre
Auger Observatory. The energy spectrum shows a clear steepening at the
energy expected for the GZK pion photoproduction by protons or
photodisintegration of iron.  The structure is very similar to the
HiRes spectrum, differing only by about a 20\% shift in energy.  The
status of the correlation of arrival directions above 55 EeV with AGN
locations was updated with new data.  The correlation has decreased
from roughly 70\% to 40\%.  However, the 2.5$\sigma$ statistical
significance of the deviation from the 21\% isotropic expectation 
is the same as when the correlation was first reported in 2007.
Figure \ref{auger_corr} shows the history and present status of the
AGN correlation.

\begin{figure}[ht]
\includegraphics[width=80mm]{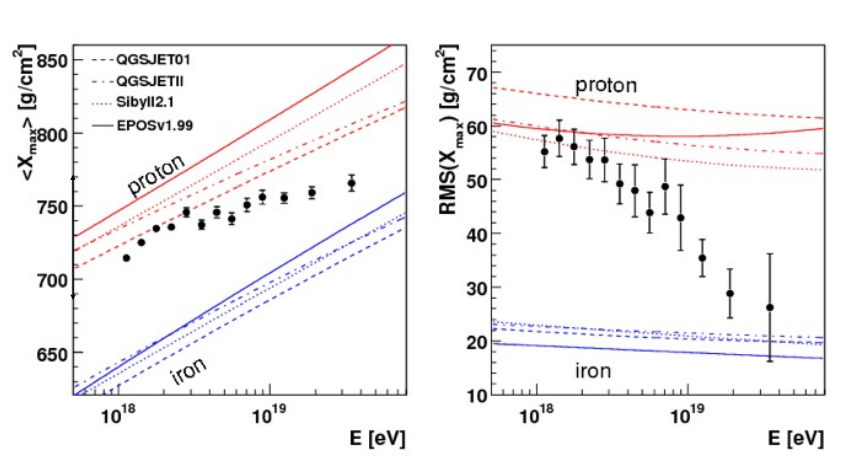}
\caption{Auger $X_{max}$ results.  The left plot shows the mean
  $X_{max}$ as a function of energy, and the right plot shows the
  width of the $X_{max}$ distribution for each energy bin.} \label{auger_xmax}
\end{figure}

\begin{figure}[h]
\includegraphics[width=80mm]{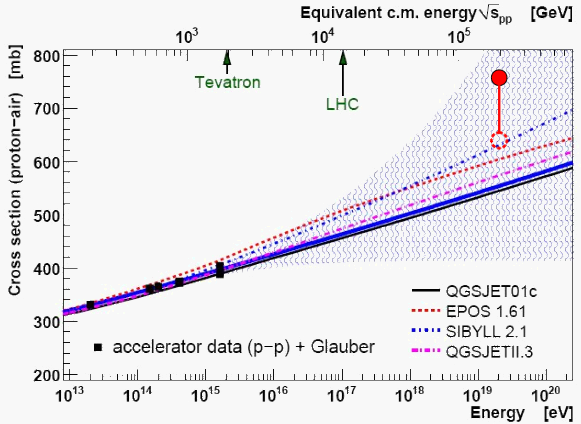}
\caption{The red dot is a conservative lower limit for the proton-air
  cross section based on the narrow distribution of $X_{max}$ values
  near 20 EeV {\it if the primary particles are protons}.  The dotted
  circle is the upper limit obtained from the ``1-sigma'' upper limit
  on that $X_{max}$ distribution width.} \label{auger_xsection}
\end{figure}

The Auger $X_{max}$ results appear different from those of HiRes, as
can be seen by comparing Figure \ref{auger_xmax} with Figure
\ref{hires_xmax}.  
Of special interest is the trend toward narrow $X_{max}$ distributions
in the decade from 4 to 40 EeV.  Using customary extrapolations of
hadronic models to these energies, the narrow distribution near 20 EeV
is not consistent with a mixture of heavy nuclei and protons, and it
is significantly narrower than expected for a pure proton composition.  It is
suggestive of a purely heavy composition and, in view of the magnetic
field of the Galaxy, not easily reconciled with the correlation of
arrival directions with AGNs at somewhat higher energy.  If the narrow
distribution is assumed to be protons at the next-to-highest energy
bin, for example, a conservative minimum proton-air cross section is
obtained by assuming that the spread in $X_{max}$ is entirely due to
the spread in first interaction depths.  Figure \ref{auger_xsection}
shows this lower limit in relation to some standard models of how the
proton-air cross section may rise with energy.

\section{Summary and discussion}

Accelerator experiments and cosmic ray observations are two arenas of
high energy physics that have been intertwined since particle
accelerators came into existence in the second half of the last
century.  The high energy frontier always belongs to cosmic rays, but
measurements with controlled conditions belong to the accelerators.
Indirect (air shower) studies of cosmic rays rely on models of
particle interactions, and presently the models extrapolate
interaction properties to energies which have not been investigated
experimentally.  For example, different extrapolations result in
different composition inferences near the knee using KASCADE data.  It
is therefore exciting that detectors like TOTEM, LHCf, and CASTOR will
be providing data in the essential forward region.  The center of mass
interaction energy is now 7 TeV and will increase to 14 TeV.  As seen
in Figure \ref{auger_xsection}, a measured cross section at the LHC
energy will be a powerful constraint on extrapolations to the regime
of ultra-high energy cosmic rays.  Air shower simulations also require
accurate modeling of secondary interactions.  The MIPP measurements,
with its variety of beam particles and nuclear targets and with its
measurements over all of phase space, will provide vital data for
this.

Cosmic ray observations are now rich in details over the entire energy
spectrum of ten decades.  Satellite and balloon missions have measured
spectra of individual elements or groups of elements almost up to the
knee.  Anisotropy has been mapped by Milagro, IceCube, Tibet AS array,
and ARGO.  KASCADE-Grande continues to refine measurements up to 100
PeV in order to determine the spectra of individual element groups in
the region of the knee and above.  Ultra-high energy cosmic ray
observatories, Auger and TA, are pushing down to 100 PeV from above.
At the same time, they are looking to the highest energy cosmic rays
for clues about the ultra-high energy extragalactic sources as well as
the nature of particle interactions above 300 TeV center-of-mass energy.
Fundamental questions about the origins and propagation of cosmic rays
persist at all energies.  The measurements have improved in
statistics and systematics.  Interpreting the air shower measurements
requires knowledge of particle interactions that must be obtained from
accelerator experiments.  These ISVHECRI meetings play an important
role.

\bigskip 

\end{document}